\documentstyle[12pt]{article}
\def\a{\begin{eqnarray}}
\def\b{\end{eqnarray}}
\def\0{\nonumber}
\begin{document}
\begin{flushright}
SISSA-ISAS 48/92/EP
\end{flushright}
\vskip0.5cm
\centerline{\LARGE An alternative approach to KP hierarchy}
\vskip0.2cm
\centerline{\LARGE in matrix models}
\vskip1.5cm
\centerline{\large  L.Bonora}
\centerline{International School for Advanced Studies (SISSA/ISAS)}
\centerline{Via Beirut 2, 34014 Trieste, Italy}
\centerline{INFN, Sezione di Trieste.  }
\vskip0.5cm
\centerline{\large C.S.Xiong}
\centerline{International School for Advanced Studies (SISSA/ISAS)}
\centerline{Via Beirut 2, 34014 Trieste, Italy}
\vskip5cm
\abstract{We show that there exists an alternative procedure in
order to extract differential hierarchies, such as the KdV
hierarchy, from one--matrix models, without taking a
continuum limit. To prove this we introduce the Toda
lattice and reformulate it in operator form. We then
consider the reduction to the systems appropriate for
one--matrix model}

\vfill
\eject

\section{Introduction}

The one Hermitean matrix model \cite{MM} (\cite{MMR} contains a few of the
numerous existing reviews) has attracted a lot of attention recently: it is
a new and powerful tool in the study of two dimensional gravity (coupled
to matter) and it has revealed an extremely interesting mathematical
structure. The latter is synthesized by the appearence of a KdV hierarchy
of which the partition function is a $\tau$--function, and by the string
equation which can be cast in the form of Virasoro constraints over the
partition function. It has been shown recently that the KdV hierarchy
characterizes both topological gravity and intersection theory of the moduli
space of Riemann surfaces \cite{W1},\cite{W2},\cite{Ko}. On the other hand the
Virasoro constraints appear naturally in topological field theory
\cite{DVV},\cite{FKN}. One matrix models are at the core of all
these developments, and although they have been the object of intensive
research, still they may hide some surprises.

In this paper we will eventually deal with the KdV hierarchy ensuing from
one--matrix models. It is well--known (see \cite{W2},\cite{BMX}) that
one--matrix models are characterized by a linear system whose
integrability conditions form a discrete hierarchy (i.e. a hierarchy
of differential--difference
equations). This turn out to be a reduced case of the so--called
Toda lattice hierarchy. What in the literature is known as the KdV
hierarchy associated to a one--matrix model is obtained as a
continuum limit of the above discrete hierarchy: it is formed by
a hierarchy of purely differential equations, among which we find
the celebrated KdV equation.

To avoid misunderstandings let us insist on the difference between
the two above--mentioned types of hierarchies.
They are typically represented by the two hierarchies (\ref{dKP}) and
and (\ref{cKP}) below. They have the same form, but in the first case $Q$ is
an (infinite) matrix and the equations can be interpreted as
differential--difference equations; for this reason we call this hierarchy
{\it discrete}. In the second $\hat Q_n$ is a pseudo--differential operator and
the equations involved are purely differential; for this reason we call
this hierarchy {\it differential}. Of the latter type is the KdV hierarchy
met in the literature.

In this letter we want to point out that taking a continuum
limit is not necessary in order to get a differential hierarchy. There
exists an alternative procedure by which we can extract a differential
hierarchy from the discrete linear system associated to one--matrix
models {\it without reference to any limiting procedure}. Said differently,
the discrete linear system contains already a differential hierarchy,
which can be reduced to the KdV hierarchy, without us being obliged
to resort to a limiting procedure -- which presumably causes some loss
of information. In this hierarchy the first flow parameter $t_1$
plays the role of space coordinate.

The general setting proposed here seems to be closer than the
usual one, based on the continuum limit, to the spirit
of Kontsevich model.

The paper is organized as follows. We first introduce the Toda lattice in
general (section 2), which involves $\infty\times \infty$ matrices,
and we show in this general case how to pass from the
matrix formulation to the (differential) operator formulation (section 3).
Then we consider reductions of the above system to semi--infinite matrices
(section 4). Finally (section 5) we consider the reduction to the linear system
appropriate for one--matrix models (the Toda chain), and, in particular,
{\it by a further reduction} we recover the KdV hierarchy.

\section{The Toda lattice hierarchy}

In this section we introduce the so-called Toda lattice hierarchy. The
main reference in this context  is the paper of Ueno and Takasaki \cite{UT}.
However we will present it in a form which is more suitable for our purposes,
i.e. mainly by means of the associated linear system.

Let us first introduce some notations.
Given a matrix $M$, we will denote by $M_-$ the strictly lower triangular part
and by $M_+$ the upper triangular part including the main diagonal.
Unless otherwise specified, we will be dealing with $\infty \times
\infty$ matrices.
As usual $E_{ij}$ will denote the matrix
$(E_{ij})_{kl}=\delta_{ik}\delta_{jl}$.
We will also use
\a
I_{\pm}\equiv \sum_{i=-\infty}^{\infty}E_{i,i\pm 1},
\quad\qquad
\rho=\sum_{i=-\infty}^{\infty}i E_{ii}\0
\b
Throughout this paper $\lambda$ denotes the spectral parameter,
and $\Lambda$ represents an infinite dimensional column vector whose
components $\Lambda_n,~n\in {\bf Z}$ are given by
\a
\Lambda_n = \lambda^n \0
\b
The vector $\Lambda$ is our elementary starting point. From it, by means
of matrix transformations, we obtain can other vectors. A useful one
is $\eta$
\a
\eta=exp\bigl({\sum_{r=1}^{\infty}t_r\lambda^r}\bigl)\Lambda \label{eta}
\b
whose components are
\a
\eta_n=exp\bigl({\sum_{r=1}^{\infty}t_r\lambda^r}\bigl)\lambda^n\0
\b
In (\ref{eta}) $t_r$ are time or flow parameters.
On $\eta$ one can naturally define a (elementary) linear system
\a
&&\lambda\eta=\partial\eta
=I_+\eta\0\\
&&\lambda^r\eta={\partial\over{\partial t_r}}\eta
=\partial^r\eta
=I^r_+\eta\label{etasystem}\\
&&\lambda^{-r}\eta
=\partial^{-r}\eta
=I^r_-\eta \0\\
&&{\partial\over{\partial \lambda}}\eta=P_0\eta,\qquad
P_0=\rho I_-+\sum_{r=1}^{\infty}rt_rI_+^{r-1}
\0
\b
Thoughout the paper $\partial$ denotes the derivative ${\partial \over
{\partial  t_1}}$. Moreover $\partial^{-1}$ denotes formal integration.
Since for $(\infty\times\infty)$ matrices,
$$
[I_+, \rho I_-]=1
$$
we see that the spectral and flow equations are
automatically compatible.

A crucial ingredient in the following construction is the (invertible)
``wave matrix" $W$:
\a
W=1+\sum_{i=1}^{\infty}w_i I_-^i= 1+\sum_{i=1}^{\infty}
\sum_{n=-\infty}^{+\infty}w_i(n)E_{n,n-i}  \label{W}
\b
So $w_i= \{w_i(n)|n\in {\bf Z}\}$ are infinite diagonal matrices,
and $w_i(n)$ are functions of the time parameters.  We impose them to be
determined by the equations of motion
\a
{\partial\over{\partial t_r}}W=Q^r_+W-WI_+^r\label{emW}
\b
where $Q$ is the infinite matrix
\a
Q=WI_+W^{-1}\label{Q}
\b

Another important object is the vector $\Psi$
\a
\Psi=W\eta\label{Psi}
\b
In terms of all these objects the dynamical system we have defined can be
written as
\a
Q\Psi&=&\lambda\Psi\label{LS1}\\
{\partial\over{\partial t_r}}\Psi&=&Q^r_+\Psi\label{LS2}\\
{\partial\over{\partial \lambda}}\Psi&=&P\Psi\label{LS3}
\b
where
\a
P= WP_0W^{-1} \label{P}
\b

The compatibility conditions of this linear
system form
the so--called {\it discrete} KP--hierarchy
\a
{\partial\over{\partial t_r}}Q=[Q^r_+, Q]\label{dKP}
\b
together with the {\it trivial} relation
\a
[Q, P]=1\label{dstring}
\b

We should perhaps recall that what we have done so far is at a purely formal
level and does not bear yet any relation to matrix models. In particular
we insist that eq.(\ref{dstring}) does not imply any constraint on the
dynamical
system.

To end this section let us make the above formulas more explicit and extract
a few relations that will be useful  in the following.
{}From the equation of motion (\ref{emW}) we consider in particular
\a
{\partial\over{\partial t_1}}w_i(n)
=w_{i+1}(n+1)-w_{i+1}(n)
-w_i(n)\bigl(w_1(n+1)-w_1(n)\bigl)\label{w'}
\b
Let us introduce a piece of terminology by saying that for any matrix its
elements in the $n-th$ row belong to the $n-th$ sector. Therefore, in regard
to eq.(\ref{w'}) we can say that the flow in the $n-th$ sector depends
only on the coordinates of the $n-th$ and $n+1-th$ sectors.

{}From (\ref{Q}) we see that
\a
Q=I_++\sum_{i=0}^{\infty}a_i I_-^i\label{ai}
\b
The $a_i$'s are new coordinates of the system, which can be uniquely expressed
in terms of the $w_i$'s. For example \footnote{These equations show that the
two sets of the variables $a_i$'s
and $w_i$'s can be obtained
from each other. However, strickly speaking,
this one--to--one correspondence is only due to the fact that we have
choosen a special form of $W$--matrix (\ref{W}).
Generally, for a given matrix $Q$,
$W$ is not uniquely determined.}
\a
a_0(n)&=&w_1(n)-w_1(n+1)\0\\
a_1(n)&=&w_1(n)\bigl(w_1(n+1)-w_1(n)\bigl)+w_2(n)-w_2(n+1)\0\\
a_2(n)&=&w_2(n)\bigl(w_1(n+1)-w_1(n)\bigl)
+w_1(n-1)\bigl(w_2(n+1)-w_2(n)\bigl)\0\\
&&+w_1(n)w_1(n-1)\bigl(w_1(n+1)-w_1(n)\bigl)
+w_3(n)-w_3(n+1)\0
\b

Another useful representation is obtained by inverting eq.(\ref{Q})
\a
I_+=W^{-1}QW=Q+\sum_{i=0}^{\infty}q_iQ^{-i}\0
\b
the $q_i$'s are another set of diagonal matrices, which can be
expressed in terms of $a_i$'s or $w_i$'s.
It is worth noting that
\a
I_-=(I_+)^{-1}=Q^{-1}+\ldots \0
\b
which results in the following equality
\a
Q^r_+\equiv Q^r-Q^r_-=Q^r+\sum_{i=1}^{\infty}q_{r,i}Q^{-i}
\qquad \forall r\geq1 \label{Q+}
\b

Finally, from (\ref{P}), we have
\a
P=\sum_{r=1}^{\infty}rt_rQ^{r-1}+\sum_{i=0}^{\infty}v_iQ^{-i-1}\label{P1}
\b
Once again, $v_i$'s are diagonal matrices, and $v_0=\rho$.{}\footnote{
Since $W\rho I_-W^{-1}=\rho Q^{-1}+[W,\rho]I_-W^{-1}$, the commutator
is a strictly lower triangular matrix, so the second part at most
contributes to the term $Q^{-2}$, which ensures that $v_o=\rho$.}

We will see later on that the string equation of matrix models can be
obtained by imposing a constraint on the coordinates $v_i$.

\section{From the discrete hierarchy to the differential hierarchy}

In the previous section we introduced the usual Toda lattice.
The discrete KP hierarchy we obtained is known as the Toda lattice hierarchy.
It consists of an infinite set of differential-difference equations.
In this section we show that passing from the matrix formalism of the previous
section to a related (pseudo--differential) operator formalism, we can obtain
a new hierarchy which consists merely of differential equations.

The operator formalism alluded to before is introduced as follows.
We recall that
equation (\ref{eta}) implies
\a
\eta_n=\partial^{n-m}\eta_m,\qquad\qquad \forall n,m:\hbox{\rm
integers}\0
\b
This leads to
\a
\Psi_n=(W\eta)_n=
\sum_{i=-\infty}^nW_{ni}\eta_i
=\sum_{i=-\infty}^nW_{ni}\partial^{i-n}\eta_n
={\hat W}_n\eta_n
\label{Psin}
\b
where we have defined
\a
{\hat W}_n=1+\sum_{i<n}W_{ni}\partial^{i-n}
=1+\sum_{i=1}^{\infty}w_i(n)\partial^{-i}
\label{Wn}
\b
This tells us that the ``wave" matrix $W$ can be considered as
an infinite column vector, whose components are differential operators.
The operator $\hat W_n$ can be inverted
\a
\eta_n={\hat W}_n^{-1}\Psi_n\0
\b
In this formalism the spectral equation (\ref{LS1}) becomes
\a
\lambda\Psi_n=\lambda{\hat W}_n\eta_n
={\hat W}_n\partial\eta_n
={\hat W}_n\partial{\hat W}_n^{-1}\Psi_n
={\hat Q}_n\Psi_n\0
\b
Here we have introduced an infinite set of KP--type differential operators
\a
{\hat Q}_n={\hat W}_n\partial {\hat W}_n^{-1}
=\partial+\sum_{i=1}^{\infty}u_i(n)\partial^{-i}\qquad\forall
n~ {\rm integer}\label{Qn}
\b
The variables $u_i$'s can also be understood as a set of coordinates
of the system. Inverting this relation we obtain
\a
\partial={\hat W}_n^{-1}{\hat Q}_n{\hat W}_n
={\hat Q}_n+\sum_{i=0}^{\infty}q_i(n){\hat Q}_n^{-i}
\qquad\forall n ~ {\rm integer}\label{Qinverse}
\b
It is easy to see that this mapping from matrices to operators
maps the upper triangular part of
a given matrix into the differential part of the
operator, and the lower triangular part of the matrix
to the formal integration part of the operator. In particular we
have
\a
\bigl(Q^r_+\Psi\bigl)_n\longrightarrow \bigl({\hat Q}_n\bigl)^r_+
\Psi_n\0
\b
For an operator, the subscript ``+" selects, as usual, the non--negative
powers of
the derivative $\partial$. Going on with the transcription of the Toda lattice
linear system in the operator formalism, we can now
rewrite the flow equations (\ref{LS2})
\a
{\partial\over{\partial t_r}}\Psi_n
&=&\bigl({\partial\over{\partial t_r}}{\hat W}_n\bigl)\eta_n
+{\hat W}_n{\partial\over{\partial t_r}}\eta_n\0\\
&=&\bigg(\lambda^r+\bigl({\partial\over{\partial t_r}}
{\hat W}_n\bigl){\hat W}_n^{-1}\bigg)\Psi_n\0\\
&=&\bigl({\hat Q}_n^r\bigl)_+\Psi_n\qquad\forall n~{\rm integer}\0
\b
Finally we can rewrite the linear system as follows
\a
&&{\hat Q}_n\Psi_n=\lambda\Psi_n\label{cLS1}\\
&&{\partial\over{\partial t_r}}\Psi_n
=\bigl({\hat Q}_n^r\bigl)_+\Psi_n, \qquad \forall n~{\rm integer}
\label{cLS2}\\
&&{\partial\over{\partial\lambda}}\Psi_n=\bigl(\sum_{r=1}^{\infty}rt_r
{\hat Q}_n^{r-1}
+\sum_{i=0}^{\infty}v_i(n){\hat Q}_n^{-i-1}\bigl)\Psi_n\label{cLS3}
\b
Their compatibility conditions are
\a
{\partial\over{\partial t_r}}{\hat Q}_n=[\bigl({\hat Q}_n\bigl)^r_+,{\hat
Q}_n]
\label{cKP}
\b
or in other coordinates
\a
{\partial\over{\partial t_r}}{\hat W}_n=\bigl({\hat Q}_n\bigl)^r_+
{\hat W}_n-{\hat W}_n\partial^r
\label{cKPW}
\b
The last two equations, like the previous ones, hold for any integer $n$.
Eq.(\ref{cKP}) or (\ref{cKPW}) specifies the {\it differential} hierarchy we
promised in the
introduction. It consists of an infinite set of differential equations: in the
LHS we have the first order derivatives with respect to the flow parameters,
in the RHS we have polynomials in the coordinates and derivatives of
coordinates with respect to $t_1$.

It is a bit inappropriate to speak of one hierarchy: we have in fact
an infinite number of hierarchies, one for each integer $n$. However these
hierarchies are not independent as they are related by the $t_1$ flow; we
will see later on that in the particular case of one-matrix model all
these hierarchies are isomorphic.

Let us remark that each equation in (\ref{cKPW}) involves only coordinates
of the same sector. This is not true for (\ref{cKP}); however by using
the $t_1$ flow one can express each equation only in terms of coordinates
of a single sector.

Finally let us comment on the relation between the discrete Toda hierarchy
and the hierarchies (\ref{cKP}). We have seen that from the discrete lattice
(with its hierarchy) through the operator formalism we can unambiguously
arrive at a new differential hierarchy. From a practical point of view this
corresponds to expressing the difference operations on the lattice in terms
of $t_1$ derivatives of the coordinates in one sector. Conversely from
the  latter hierarchy we can reconstruct the discrete one. Let us sketch the
argument. One starts from the linear system specified by
\a
&&L(x,t)= \partial+ u_1(x,t)\partial^{-1} + u_2(x,t)\partial^{-2}+\ldots,
\quad\quad \partial= {d\over {dx}}\0\\
&&L(x,t) \Psi(x,t,\lambda) = \lambda \Psi(x,t,\lambda)\0\\
&&{\partial\over {\partial t_r}}\Psi(x,t,\lambda)= L_+^r \Psi(x,t,\lambda)\0
\b
where $t$ represent the collection of flow parameters $t_1,t_2,\ldots$, and
the associated KP hierarchy. From the last equation we see that it is
consistent  to identify $x$ with $t_1$.
Next we can define the operator
\a
W= 1 +w_1\partial^{-1} + w_2 \partial^{-2}+ \ldots\label{aux}
\b
such that
\a
L= W\partial W^{-1} \0
\b
Now we call the $w_i$ in eq.(\ref{aux}) as $w_i(n)$, and we use the $t_1$ flow
to define the $w_i(n+1)$ as in eq.(\ref{w'}). In this way we can reconstruct
the
discrete linear system and the discrete KP hierarchy.

\section{Reduction: Semi--infinite matrices}

Let us study now the problem of reducing the general system defined
in the two previous sections to a simpler one.
In the next section we will consider a further reduction, i.e. to
the Toda chain which is relevant for one-matrix models.
We notice, first of all, that for the linear systems involved in matrix models
$\Psi$ does not contain
negative powers of $\lambda$. Therefore the Jacobi matrix $Q$
must be semi--infinite,
\a
w_i(n)=0,\qquad\qquad n<i\0
\b
This is the reduction we will study in this section.

In this reduced system for any positive integer $n$ we have an invertible
operator
with a finite number of terms
\a
{\hat W}_n=1+\sum_{i=1}^n w_i(n)\partial^{-i}\0
\b
The KP--type operator is
\a
{\hat Q}_n={\hat W}_n\partial{\hat W}_n^{-1}\0
\b
which still contains infinite many terms.

We remark here that we are still formally using $\infty \times \infty$ matrices
in order to be able to fully exploit the formalism introduced in the previous
sections. However three quadrants of these matrices become irrelevant.

So far we have been using mostly $w_i$ coordinates, but henceforth it will
be more convenient to shift to $a_i$ coordinates. We recall that they are
defined in the following way through
the spectral equation
\a
\lambda\Psi_n=\Psi_{n+1}
+\sum_{i=0}^na_i(n)\Psi_{n-i}\label{spectral}
\b
Next we want to express the RHS of this equation in terms of $\Psi_n$ only
(this is an example of the procedure outlined in the last section).
To this end we use the first flow equation
\a
\partial\Psi_n=\bigl(Q_+\Psi\bigl)_n=\Psi_{n+1}+a_0(n)\Psi_n\0
\b
or equivalently
\a
\Psi_{n+1}=\bigl(\partial-a_0(n)\bigl)\Psi_n\label{t1flow}
\b
Inverting this relation, we get
\a
\Psi_n={\hat B}_n\Psi_{n+1},
\qquad
{\hat B}_n=\partial^{-1}\sum_{l=0}^{\infty}\bigl(a_0(n)
\partial^{-1}\bigl)^l\label{B}
\b
Using this relation repeatedly
we can express any $\Psi_i(i<n)$
in terms of $\Psi_n$, i.e.
\a
\Psi_{n-r}={\hat B}_{n-r}{\hat B}_{n-r+1}\ldots{\hat B}_{n-1}\Psi_n\0
\b
Therefore the spectral equation (\ref{spectral}) can be rewritten as
\a
{\hat Q}_n\Psi_n
=\bigg(\partial+\sum_{i=1}^na_i(n){\hat B}_{n-i}{\hat B}_{n-i+1}
\ldots{\hat B}_{n-1}\bigg)\Psi_n\label{spectral'}
\b
and the n--th KP--type operator becomes
\a
{\hat Q}_n=\partial+\sum_{i=1}^na_i(n){\hat B}_{n-i}{\hat B}_{n-i+1}
\ldots{\hat B}_{n-1}\label{Q'}
\b

{}From the first KP--equation in matrix form
\a
{\partial\over{\partial t_1}}Q=[Q_+, Q]\0
\b
written in terms of coordinates
\a
{\partial\over{\partial t_1}}a_i(n)=a_i(n)^{'}
=a_{i+1}(n+1)-a_{i+1}(n)+a_i(n)\bigl(a_0(n)-a_{0}(n-i)\bigl)\0
\b
which just makes the connection between
the coordinates $a_i(n)(i\leq n)$'s
in the n--th sector and
the coordinates $a_i(j)$'s
in the j--th sectors, $j\leq n+1$.
We conjecture that, using these relations, one should be able to write
\a
{\hat Q}_n=\partial+\sum_{l=1}u_l(n)\partial^{-l}\0
\b
where the functions $u_l(n)$'s only depend on the coordinates
in the n--th sector. We will explicitly show this property below in the
case of one-matrix model.
The equations of motion (the remaining $t_r(r\geq2)$--flows), as expected,
take the form of the differential KP--hierarchy
\a
{\partial\over{\partial t_k}}{\hat Q}_n=[\bigl({\hat Q}_n\bigl)^k_+,
{\hat Q}_n]\label{cKP1}
\b

Let us now discuss the integrability of the reduced system.
We construct a bi--Hamiltonian structure as follows.
Define an inner product in the differential operator space as usual
\a
Tr(A)=\int\/dx A_{-1}(x),\qquad\qquad
A=\ldots+A_{-1}(x)\partial^{-1}+\ldots\0
\b
Using this trace operation, we can write a functional of the
$a_j(x)$'s in the following way
\a
f_X({\hat Q}_j)=Tr({\hat Q}_jX)\equiv{\hat Q}_j(X)\0
\b
where $X$ is a pure differential operator.
Then the coadjoint analysis shows us that there exist
two compatible Poisson brackets
\a
\{f_X, f_Y\}_1({\hat Q}_n)&=&{\hat Q}_n([Y, X])\label{PB1}\\
\{f_X, f_Y\}_2({\hat Q}_n)&=&<(X{\hat Q}_n)_+Y{\hat Q}_n>
-<({\hat Q}_nX)_+{\hat Q}_nY>
\label{PB2}
\b
The relevant conserved quantities (in involution) are
\a
H_k={1\over k}Tr({\hat Q}_n^k), \qquad\qquad \forall k\geq1\0
\b
and the compatibility reads
\a
\{H_{r+1},f\}_1=\{H_r, f\}_2\qquad\hbox{\rm for any function}\quad f
\b

\section{Toda chain and one-matrix models}

The case relevant to one--matrix models is specified by the conditions
\a
a_0(j)=S_j,\qquad
a_1(j)=R_j,\qquad
a_i(j)=0,\qquad\forall i\geq2\0
\b
The first (i.e. $t_1$) flow equation is
\a
{\partial\over{\partial t_1}}S_j&=&S_j^{'}=R_{j+1}-R_j\label{Sj}\\
{\partial\over{\partial t_1}}R_j&=&R_j^{'}=R_j(S_j-S_{j-1})\label{Rj}
\b
{}From the second equality we have
\a
S_{j-1}=S_j-{{R_j^{'}}\over{R_j}}\0
\b
Therefore the j--th KP--type operator is
\a
{\hat Q}_j
=\partial+R_j{\hat B}_{j-1}
\equiv \partial+\sum_{l=1}u_l(j)\partial^{-l}\label{Qj}
\b
The first few $u_l(j)$'s are
\a
&&u_1(j)=R_j,\qquad\qquad u_2(j)=-R_j^{'}+R_jS_j\0\\
&&u_3(j)=R_j^{''}-2R_j^{'}S_j-R_jS_j^{'}+R_jS_j^2\0\\
&&u_4(j)=-R_j^{'''}+3R_j^{''}S_j+3R_j^{'}S_j^{'}-3R_j^{'}S_j^2
-3R_jS_jS_j^{'}
+R_jS_j^{''}+R_jS_j^3\0\\
&&u_5(j)=R_j^{''''}-4R_j^{'''}S_j-6R_j^{''}S_j^{'}+6R_j^{''}S_j^2
-4R_j^{'}S_j^{''}-4R_j^{'}S_j^3\0\\
&&~~~~~~~~+12R_j^{'}S_jS_j^{'}+4R_jS_jS_j^{''}
+3R_j{S_j^{'}}^2\0\\
&&~~~~~~~~-6R_jS_j^2S_j^{'}-R_jS_j^{'''}
+R_jS_j^4\0
\b
As anticipated in the previous section the $u_l$'s depend only on the
coordinates of one sector. The consequence of this is that
the infinite many KP--hierarchies we discussed about at the end of section 3
are all isomorphic.

Since the following analysis is universal, i.e. is the same for all sectors,
we simply omit the subscript ``j", and denote $R_j(t_1)$, $S_j(t_1)$
by $R(t_1)$ and $S(t_1)$, respectively.

{}From the above general discussion,
we can derive two compatible Poisson brackets, which are
\a
\{R(x), R(y)\}_1&=&0\0\\
\{S(x), S(y)\}_1&=&0\0\\
\{R(x), S(y)\}_1&=&-\partial_x\delta(x-y)\0
\b
and
\a
\{R(x), R(y)\}_2&=&-\bigl(2R(x)\partial_x+R^{'}(x)\bigl)\delta(x-y)\0\\
\{S(x), S(y)\}_2&=&-2\partial_x\delta(x-y)\0\\
\{R(x), S(y)\}_2&=&\bigl(\partial^2_x-S(x)\partial_x\bigl)\delta(x-y)\0
\b

We are now going to write down the flow equations of this system.
To do this one can directly use equation (\ref{cKP1}).
A more concise way is to introduce the two polynomials
\a
F_r(x)={{\delta H_r}\over{\delta S(x)}}
\qquad
G_r(x)={{\delta H_r}\over{\delta R(x)}}
\qquad\forall r\geq1\0
\b
In particular
$$
F_1=0,\qquad\qquad G_1=1
$$
The compatibility condition of two Poisson brackets shows the existence
of the following recursion relations among these polynomials
\a
F^{'}_{r+1}&=&SF^{'}_r-F^{''}_r+2RG^{'}_r+R^{'}G_r\label{F}\\
G^{'}_{r+1}&=&2F^{'}_r+G^{''}_r+\bigl(SG_r\bigl)^{'}\label{G}
\b
In terms of these polynomials, the general equations of
motion can be written as
\a
{\partial\over{\partial t_r}}S&=&G^{'}_{r+1}\0\\
{\partial\over{\partial t_r}}R&=&F^{'}_{r+1}\0
\b
For example, the first two flows have the following explicit forms
\a
{\partial\over{\partial t_2}}S&=&
S^{''}+2SS^{'}+2R^{'}\0\\
{\partial\over{\partial t_2}}R&=&-R^{''}+2(RS)^{'}\label{R2}
\b
and
\a
{\partial\over{\partial t_3}}S&=&
S^{'''}+3SS^{''}+3{S^{'}}^2+6(RS)^{'}+3S^2S^{'}\0\\
{\partial\over{\partial t_3}}R&=&R^{'''}-3R^{''}S
+6RR^{'}+3(RS^2)^{'}-3R^{'}S^{'}\label{R3}
\b

This is the hierarchy characterizing the Toda chain and one-matrix models.

Hereafter we show that by further restricting the system we can obtain
the KdV hierarchy.
This is achieved in the following way. Let us set
\a
S=0\0
\b
and let  us discard the $t_{2r}$ flows.
Then from eqs.(\ref{F},\ref{G}) we get
\a
G_{2r}&=&0\0\\
F'_{2r+2}&=& (\partial^3 +4 R \partial +2R')F_{2r}\0
\b
utilizing the initial condition $F_2=R$.
This is the recursion relation for KdV hierarchy.
In particular, as can be seen also from (\ref{R3}), we have
\a
{\partial\over{\partial t_3}}R=R^{'''}
+6RR^{'}\label{KdV}
\b
The fact that in one--matrix models the conditions $S_j=0$ and $t_{2r+1}=0$
are related does not prevent us from drawing the above conclusions.
Here we have been considering the reduction of the system ensuing from
a one--matrix model under the condition $S=0$, and this is of course
legitimate. Considered in the context of one--matrix model it seems to
correspond to a rather singular situation. This is perhaps not entirely
surprising since the KdV hierarchy corresponds to a topological point.
On the other hand also in the continuum
limit approach the reduction to the even potentials presents not fully
understood aspects (see, for example (\cite{BMX}).
We think an understanding
of this problem will be facilitated by the analysis of the full
hierarchy (\ref{R2},\ref{R3}). We will return to this point elsewhere.

What we said so far in this section is valid for one--matrix models but
does not characterize them completely. As is well known we have to impose
the string equation.
As we already noticed, the equation
\a
\relax [Q, P]=1\label{QP}
\b
does not imply any restriction on the model as long as $P$ has the general
form (\ref{P}). One-matrix models are characterized by the following
restriction on the form of $P$
\a
P_-=0,\quad\quad {\rm i.e.}\quad P= \sum_{r=1}^\infty rt_r Q^{r-1}_+
\label{1mm}
\b
By string equation we mean (\ref{QP}) together with the
condition (\ref{1mm}). Starting from it we can set out for the analysis of
the critical points.

To end this section let us discuss the Virasoro constraints for the
system corresponding to one--matrix models.
Since they were already derived along similar lines in a previous paper
\cite{BMX}, we will simply sketch the derivation here.
{}From the string equation we can derive
\a
&&d_{-1} R=0\0\\
&& d_{-1} S+1=0\label{dS}
\b
where
\a
d_{-1}= \sum_{r=2}^\infty rt_r {\partial\over{\partial t_{r-1}}}\0
\b
Eqs.(\ref{dS}) can be written in the form
\a
\sum_{r=2}^\infty rt_r F_r' &=&0 \0\\
\sum_{r=2}^\infty rt_r G_r' +1 &=&0 \0
\b
Using the recursion relations (\ref{F},\ref{G}) we can obtain other similar
equations, for example
\a
\sum_{r=1}^\infty rt_r F_{r+1}'+2R &=&0 \0\\
\sum_{r=1}^\infty rt_r G_{r+1}' +S &=&0 \0
\b
etc.

Now using the definition
\a
{\partial\over{\partial t_r}}\ln Z_N(t)=\sum_{i=0}^{N-1}Q^r_{ii}\0
\b
we can recast the above equations in the familiar form
of the Virasoro constraints
\a
&&\Big(\sum_{r=2}^\infty rt_r {\partial\over{\partial t_{r-1}}}+Nt_1\Big) Z_N
=0\0\\
&&\Big(\sum_{r=1}^\infty rt_r {\partial\over{\partial t_{r}}}+N^2\Big)Z_N=0\0
\b

\section{Discussion}

Starting from the Toda lattice in the traditional form, we have extracted
a continuous linear system (without taking a continuum limit).
By reducing it we have been able to show that
the discrete system corresponding to one--matrix models gives rise
naturally to a differential hierarchy and in particular to the KdV
hierarchy.

A few differences with the more common limiting treatment in one-matrix
models should be stressed:

-- the space parameter in our approach is $t_1$ (not $t_0$, which does not
show up here);

-- the Virasoro constraints we found in the previous section
are the same as the Virasoro constraints of the discrete system \cite{BMX},
and therefore differ from the Virasoro constraints one finds after taking the
continuum limit;

-- the complete differential hierarchy corresponding to one--matrix
models does not seem to coincide exactly with any of the continuous
hierarchies proposed so far \cite{P}; but further research is needed
on this point.

{\bf Acknowledgements} One of us (L.B.) would like to thank the Instituto
de Fisica Teorica -- UNESP for the kind hospitality extended to him
during the completion of this paper.

\end{document}